\documentclass[prd,showpacs,preprintnumbers,twocolumn,amsmath,nofootinbib,amssymb]{revtex4}
\usepackage{graphicx,color,dcolumn,booktabs,bm}
\usepackage{longtable,lscape}
\usepackage{txfonts}
\usepackage{overpic}
\usepackage{amssymb}
\usepackage{epstopdf}
\usepackage{indentfirst}
\usepackage{feynmf}   
\usepackage{slashed}  
\usepackage{cases}
\usepackage{color}
\usepackage{float}
\usepackage{multirow}
\usepackage{graphicx,color,dcolumn,booktabs,bm}
\usepackage{epsfig,dsfont,amssymb,amsmath,amsfonts,amsbsy,mathrsfs}

\graphicspath{{Figures/}} %

\usepackage{hyperref}
\hypersetup{colorlinks,citecolor=blue,anchorcolor=red,menucolor=red, linkcolor=red,filecolor=red,runcolor=red,urlcolor=blue,frenchlinks=red}

\makeatletter
\@addtoreset{equation}{section}
\makeatother

\allowdisplaybreaks

\begin{document}

\title{Strong decays of the newly $P_{cs}(4459)$ as a strange hidden-charm $\Xi_c\bar{D}^*$ molecule}

\author{Rui Chen$^{1,2}$}
\email{chen$_$rui@pku.edu.cn}
\affiliation{
$^1$Center of High Energy Physics, Peking University, Beijing
100871, China\\
$^2$ School of Physics and State Key Laboratory of Nuclear Physics and Technology, Peking University, Beijing 100871, China}

\begin{abstract}
  In our former work [arXiv:2011.07214], the $P_{cs}(4459)$ observed by the LHCb Collaboration can be explained as a coupled strange hidden-charm $\Xi_c\bar{D}^*/\Xi_c^*\bar{D}/\Xi_c'\bar{D}^*/\Xi_c^*\bar{D}^*$ molecule with $I(J^P)=0(3/2^-)$. Here, we further discuss the two-body strong decay behaviors of the $P_{cs}(4459)$ in the meson-baryon molecular scenario by input the former obtained bound solutions. Our results support the $P_{cs}(4459)$ as the strange hidden-charm $\Xi_c\bar{D}^*$ molecule with $I(J^P)=0(3/2^-)$. The relative decay ratio between $\Lambda_cD_s^*$ and $J/\psi\Lambda$ is around 10, where the partial decay width for the $\Lambda_cD_s^*$ channel is around 0.6 to 2.0 MeV.
\end{abstract}


\maketitle

\section{introduction}

In 2019, the LHCb Collaboration discovered three narrow hidden-charm pentaquarks, namely $P_c(4312)$, $P_c(4440)$, and $P_c(4457)$, by using the combined data set collected in Run 1 plus Run 2 \cite{Aaij:2019vzc}. These three $P_c$ states locate just below the $\Sigma_c\bar{D}^{(*)}$ continuum thresholds, they are very likely to be $\Sigma_c\bar{D}^{(*)}$ hidden-charm molecular pentaquarks. Several phenomenological models have been adopted to calculate the mass spectrum of the meson-baryon hidden-charm molecules, like the QCD sum rule, the meson-exchange model, the quark delocalization model, and so on (see review papers \cite{Chen:2016qju,Liu:2019zoy,Brambilla:2019esw,Guo:2017jvc,Esposito:2016noz,Hosaka:2016pey} for more details).
In particular, through adopting the one-boson-exchange model (OBE) and considering the coupled channel effect, we have demonstrated that the $P_c(4312)$, $P_c(4440)$, and $P_c(4457)$ are corresponding to the loosely bound $\Sigma_c\bar{D}$ state with $I(J^P)=1/2(1/2^-)$, $\Sigma_c\bar{D}^*$ state with $I(J^P)=1/2(1/2^-)$, and $\Sigma_c\bar{D}^*$ state with $I(J^P)=1/2(3/2^-)$, respectively \cite{Chen:2019asm}. And the coupled-channel effect plays an important role in generating hidden-charm molecular pentaquarks.

The hadronic molecule is an important component of exotic states. Experimental and theoretical studies on the hadronic molecules can deepen our understanding of the nonperturbative behavior of quantum chromodynamics (QCD). Especially, not only the study of the mass spectrums but also the predictions of the decay behaviors for the $P_c$ states can help us to test the binding mechanism of pentaquark states. So far, many groups have discussed the strong decay behaviors of the $P_c$ states in the meson-baryon hadronic molecular picture. For example, the decay branch fraction of $P_c\to \eta_c p$ and $P_c\to J/\psi p$ processes were predicted by using the heavy quark symmetry \cite{Wang:2019spc,Voloshin:2019aut,Sakai:2019qph,Chen:2020pac,Xu:2019zme,Wang:2019hyc,Gutsche:2019mkg}. The effective Lagrangian method was adopted to study the partial widths of all the allowed decay channels for these $P_c$ states at the hadronic level \cite{Lin:2019qiv}. As we seen, all the results are model dependent. The coupled channel effect is not well taken into consideration in the strong decay of the $P_c$ states.

Recently, the LHCb Collaboration reported an evidence of a possible strange hidden-charm pentaquark $P_{cs}(4459)$ decaying to the $J/\psi\Lambda$ channel in the $\Xi_b^-\to J/\psi \Lambda K^-$ process \cite{1837464}. The smallest significance is 3.1 $\sigma$. Its resonance parameters are $M=4458.8\pm2.9_{-1.2}^{+4.7}~\text{MeV}$, $\Gamma= 17.3\pm6.5_{-5.7}^{+8.0}~\text{MeV}$, respectively. The spin-parity $J^P$ of the $P_{cs}(4459)$ remains undetermined due to the lacking of the experimental data.

After the observation of the $P_{cs}(4459)$, it was interpreted as the $\Xi_c\bar{D}^*$ strange hidden-charm molecular pentaquark with $J^P=1/2^- (3/2^-)$ or the tightly pentaquark state with $J^P=1/2^-$ \cite{Chen:2020kco,Chen:2020uif,Peng:2020hql,Wang:2020eep}. In particular, we can reproduce the mass of the $P_{cs}(4459)$ in the coupled $\Xi_c\bar{D}^*/\Xi_c^*\bar{D}/\Xi_c^{\prime}\bar{D}^*/\Xi_c^*\bar{D}^*$ system with $I(J^P)=0(3/2^-)$ by adopting the one-boson-exchange model \cite{Chen:2020kco}. And the $\Xi_c\bar{D}^*$ and $\Xi_c^*\bar{D}$ systems are dominant.
In fact, the hidden-charm pentaquarks with strangeness $P_{cs}$ were predicted \cite{Wang:2019nvm,Chen:2016ryt,Anisovich:2015zqa,Wang:2015wsa,Feijoo:2015kts,
Lu:2016roh,Xiao:2019gjd,Zhang:2020cdi,Shen:2020gpw,Ferretti:2020ewe}, and suggested to search for in the $\Lambda_b(\Xi_b)\to J/\psi\Lambda K(\eta)$ \cite{Lu:2016roh,Feijoo:2015kts,Chen:2015sxa}.

In this work, we will study the two-body strong decay properties for the $P_{cs}(4459)$ as a strange hidden-charm molecule. In our calculation, we consider the coupled channel effect and input the bound state wave functions obtained in our former work \cite{Chen:2020kco}. In fact, Zou {\it et al.} already have predicted the two-body strong decay behaviors of the possible $\Lambda_{c\bar{c}}$ states in the single hadronic molecule pictures \cite{Shen:2019evi}. The obtained total widths and decay patterns can be valuable in identify the molecular assumptions and spin parities of the strange hidden-charm molecular pentaquarks.

This paper is organized as follows. After the introduction, we present the two-body strong decay amplitudes for the $P_{cs}(4459)$ as a strange hidden-charm $\Xi_c\bar{D}^*$ molecule in Sec. \ref{sec2}. The corresponding numerical results for the decay widths is given in Sec.~\ref{sec3}. The paper ends up with a summary.

\section{Two-body strong decay}\label{sec2}

For the decay process $P_{cs}\to f_1+f_2$, its decay width reads
\begin{eqnarray}
d\Gamma &=& \frac{1}{2J+1}\frac{|\bm{p}|}{32\pi^2 m_{P_{cs}}^2}|\mathcal{M}(P_{cs}\to f_1+f_2)|^2d\Omega,
\end{eqnarray}
which is expressed in the rest frame of the $P_{cs}$ state. $m_{P_{cs}}$, $J$, and $\bm{p}$ stand for the mass and spin of the initial $P_{cs}$ state and the momentum of the final state $(f_1, ~f_2)$, respectively,
\begin{eqnarray*}
|\bm{p}| &=& \sqrt{\left(m_{P_{cs}}^2-(m_{f_1}+m_{f_2})^2\right)\left(m_{P_{cs}}^2-(m_{f_1}-m_{f_2})^2\right)}/(2m_{P_{cs}}).
\end{eqnarray*}

As we discussed in Ref. \cite{Chen:2020kco}, the $P_{cs}(4459)$ can be explained as the $\Xi_c\bar{D}^*$ molecular state with $I(J^P)=0(3/2^-)$. When the binding energy is taken as $-19.28$ MeV, the probabilities for the $\Xi_c\bar{D}^*$, $\Xi_c^*\bar{D}$, $\Xi_c^{\prime}\bar{D}^*$, and $\Xi_c^*\bar{D}^*$ channels are 38.95\%, 34.58\%, 6.61\%, and 18.86\%, respectively. After introducing the coupled channel effect, the interaction for the $P_{cs}(4459)\to f_1+f_2$ process can be express as
\begin{eqnarray}\label{amplitude}
\langle f_1+f_2|P_{cs}\rangle &=& \langle f_1+f_2|\left(|\Xi_c\bar{D}^*\rangle\langle\Xi_c\bar{D}^*|
      +|\Xi_c^*\bar{D}\rangle\langle\Xi_c^*\bar{D}|\right.\nonumber\\
      &&\left.+|\Xi_c^{\prime}\bar{D}^*\rangle\langle\Xi_c^{\prime}\bar{D}^*|
      +|\Xi_c^*\bar{D}^*\rangle\langle\Xi_c^*\bar{D}^*|\right)P_{cs}\rangle\nonumber\\
 &=& \int\frac{d^3k}{(2\pi)^3}d^3re^{-i\bm{k}\cdot\bm{r}}\psi_{\Xi_c\bar{D}^*}(\bm{r})\langle f_1+f_2|\Xi_c\bar{D}^*\rangle\nonumber\\
     &&+\int\frac{d^3k}{(2\pi)^3}d^3re^{-i\bm{k}\cdot\bm{r}}\psi_{\Xi_c^{*}\bar{D}}(\bm{r})\langle f_1+f_2|\Xi_c^{*}\bar{D}\rangle\nonumber\\
     &&+\int\frac{d^3k}{(2\pi)^3}d^3re^{-i\bm{k}\cdot\bm{r}}\psi_{\Xi_c^{\prime}\bar{D}^*}(\bm{r})\langle f_1+f_2|\Xi_c^{\prime}\bar{D}^*\rangle\nonumber\\
     &&+\int\frac{d^3k}{(2\pi)^3}d^3re^{-i\bm{k}\cdot\bm{r}}\psi_{\Xi_c^{*}\bar{D}^*}(\bm{r})\langle f_1+f_2|\Xi_c^{*}\bar{D}^*\rangle.\nonumber\\
\end{eqnarray}
Here, $\psi_{\Xi_c\bar{D}^*}(\bm{r})$, $\psi_{\Xi_c^{*}\bar{D}}(\bm{r})$, $\psi_{\Xi_c^{\prime}\bar{D}^*}(\bm{r})$, and $\psi_{\Xi_c^{*}\bar{D}^*}(\bm{r})$ are the wave functions for the $\Xi_c\bar{D}^*$, $\Xi_c^{*}\bar{D}$, $\Xi_c^{\prime}\bar{D}^*$, and $\Xi_c^{*}\bar{D}^*$ channels in the $r-$coordinate space, respectively. And we define
\begin{eqnarray}
&&\langle f_1+f_2|P_{cs}\rangle = -\frac{\mathcal{M}(P_{cs}\to f_1+f_2)}{(2\pi)^{3/2}\sqrt{2E_{P_{cs}}}\sqrt{2E_{f_1}}\sqrt{2E_{f_2}}},\\
&&\langle f_1+f_2|\Xi_c^{',*}\bar{D}^{(*)}\rangle = \nonumber\\
&&\quad\quad\quad\,-\frac{\mathcal{M}\left(\Xi_c^{',*}(\bm{k})+\bar{D}^{(*)}(-\bm{k})\to f_1(\bm{p})+f_2(-\bm{p})\right)}{(2\pi)^{3/2}\sqrt{2E_{\Xi_c^{',*}}}\sqrt{2E_{\bar{D}^{(*)}}}\sqrt{2E_{f_1}}\sqrt{2E_{f_2}}}.\quad\,\,
\end{eqnarray}

There are three kinds of two-body strong decay processes, the hidden-charm modes, the open-charm modes, and the $c\bar{c}-$annihilation modes. In Table \ref{exchange}, we collect the possible two-body strong decay channels.

\begin{figure*}[!htbp]
\center
\includegraphics[width=6in]{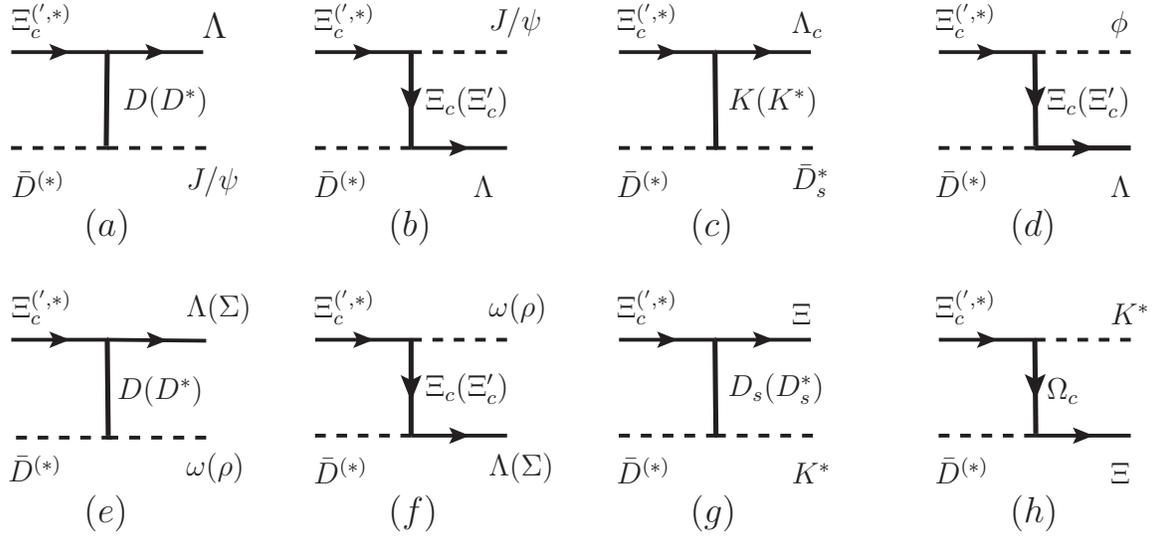}
\caption{Two-body strong decay diagrams for the $P_{cs}(4459)$ as a coupled $\Xi_c\bar{D}^{*}/\Xi_c^*\bar{D}/\Xi_c'\bar{D}^*/\Xi_c^*\bar{D}^*$ molecule with $I(J^P)=0(3/2^-)$ in the hadron level.}\label{diagram}
\end{figure*}

\renewcommand\tabcolsep{0.3cm}
\renewcommand{\arraystretch}{1.7}
\begin{table}[!htbp]
\caption{Two-body strong decay final states for the $P_{cs}(4459)$ as a $\Xi_c\bar{D}^{*}$ molecule with $I(J^P)=0(3/2^-)$. Here, the masses for the final states are in the unite of MeV. The $S$ and $D$ stand for the $S-$wave and $D-$wave decay modes, respectively.\label{exchange}}
\begin{tabular}{cccc}
\toprule[1pt]
$\eta_c\Lambda (4100)$    &$J/\psi\Lambda(4212)$    &$\Lambda_c\bar{D}_s(4254)$    &$\Lambda_c\bar{D}_s^*(4398)$\\
  $D$    &$S$   &$D$    &$S$ \\\hline
$\Sigma_c\bar{D}_s(4423)$    &$\Xi_c\bar{D}(4337)$    &$\Xi_c^{\prime}\bar{D}(4447)$    &$\phi\Lambda(2135)$\\
  $D$    &$D$   &$D$    &$S$ \\\hline
$\eta\Lambda(1662)$    &$\omega\Lambda(2007)$    &$\rho\Sigma(1971)$    &$\pi\Sigma(1326)$\\
  $D$    &$S$   &$S$    &$D$ \\\hline
$\bar{K}N(1432)$       &$\bar{K}^*N(1830)$       &$K\Xi(1810)$       &$K^*\Xi(2210)$\\
  $D$    &$S$   &$D$    &$S$ \\
 \bottomrule[1pt]
\end{tabular}
\end{table}
Because the $D-$wave interactions are strongly suppressed in comparison with the $S-$wave interactions, in the following, we only focus on the $J/\psi\Lambda$, $\Lambda_c\bar{D}_s^*$, $\phi\Lambda$, $\omega\Lambda$, $\rho\Sigma$, $\bar{K}^*N$, and $K^*\Xi$ decay channels. Figure \ref{diagram} shows the corresponding decay processes. For the isoscalar $P_{cs}$ state as the coupled $\Xi_c\bar{D}^{*}/\Xi_c^{*}\bar{D}/\Xi_c^{'}\bar{D}^{*}/\Xi_c^{*}\bar{D}^{*}$ molecule, we can obtain
\begin{eqnarray*}
\mathcal{M}_{P_{cs}\to J/\psi\Lambda} &=& \frac{1}{\sqrt{2}}\left(\mathcal{M}_{\Xi_c^{(',*)+}{D}^{(*)-}\to J/\psi\Lambda}
       -\mathcal{M}_{\Xi_c^{(',*)0}\bar{D}^{(*)0}\to J/\psi\Lambda}\right),\\
\mathcal{M}_{P_{cs}\to \Lambda_c\bar{D}_s^*} &=& \frac{1}{\sqrt{2}}\left(\mathcal{M}_{\Xi_c^{(',*)+}{D}^{(*)-}\to \Lambda_c^+D_s^{*-}}
       -\mathcal{M}_{\Xi_c^{(',*)0}\bar{D}^{(*)0}\to \Lambda_c^+D_s^{*-}}\right),\\
\mathcal{M}_{P_{cs}\to \phi\Lambda} &=& \frac{1}{\sqrt{2}}\left(\mathcal{M}_{\Xi_c^{(',*)+}{D}^{(*)-}\to \phi\Lambda^0}
       -\mathcal{M}_{\Xi_c^{(',*)0}\bar{D}^{(*)0}\to \phi\Lambda^0}\right),\\
\mathcal{M}_{P_{cs}\to \omega\Lambda} &=& \frac{1}{\sqrt{2}}\left(\mathcal{M}_{\Xi_c^{(',*)+}{D}^{(*)-}\to \omega\Lambda^0}
       -\mathcal{M}_{\Xi_c^{(',*)0}\bar{D}^{(*)0}\to \omega\Lambda^0}\right),\\
\mathcal{M}_{P_{cs}\to \rho\Sigma} &=& \frac{1}{\sqrt{6}}\left(\mathcal{M}_{\Xi_c^{(',*)+}{D}^{(*)-}\to \rho^+\Sigma^-}
       -\mathcal{M}_{\Xi_c^{(',*)0}\bar{D}^{(*)0}\to \rho^+\Sigma^-}\right.\nonumber\\
       &&\left.-\mathcal{M}_{\Xi_c^{(',*)+}{D}^{(*)-}\to \rho^0\Sigma^0}
       +\mathcal{M}_{\Xi_c^{(',*)0}\bar{D}^{(*)0}\to \rho^0\Sigma^0}\right.\nonumber\\
       &&\left.+\mathcal{M}_{\Xi_c^{(',*)+}{D}^{(*)-}\to \rho^-\Sigma^+}
       -\mathcal{M}_{\Xi_c^{(',*)0}\bar{D}^{(*)0}\to \rho^-\Sigma^+}\right),\\
\mathcal{M}_{P_{cs}\to \bar{K}^*N} &=& \frac{1}{2}\left(\mathcal{M}_{\Xi_c^{(',*)+}{D}^{(*)-}\to \bar{K}^{*0}n}
       -\mathcal{M}_{\Xi_c^{(',*)0}\bar{D}^{(*)0}\to \bar{K}^{*0}n}\right.\nonumber\\
       &&\left.+\mathcal{M}_{\Xi_c^{(',*)+}{D}^{(*)-}\to K^{*-}p}
       -\mathcal{M}_{\Xi_c^{(',*)0}\bar{D}^{(*)0}\to K^{*-}p}\right),\\
\mathcal{M}_{P_{cs}\to {K}^*\Xi} &=& \frac{1}{2}\left(\mathcal{M}_{\Xi_c^{(',*)+}{D}^{(*)-}\to {K}^{*+}\Xi^-}
       -\mathcal{M}_{\Xi_c^{(',*)0}\bar{D}^{(*)0}\to {K}^{*+}\Xi^-}\right.\nonumber\\
       &&\left.-\mathcal{M}_{\Xi_c^{(',*)+}{D}^{(*)-}\to K^{*0}\Xi}
       +\mathcal{M}_{\Xi_c^{(',*)0}\bar{D}^{(*)0}\to K^{*0}\Xi}\right).
\end{eqnarray*}
Here, we need to mention that the sum of the decay amplitude for the $\Xi_c^{(',*)}\bar{D}^{(*)}\to \bar{K}^*N$ by exchanging the $\Lambda_c/\Sigma_c$ is zero in the isospin conservation case.

\begin{figure}[!htbp]
\center
\includegraphics[width=3.in]{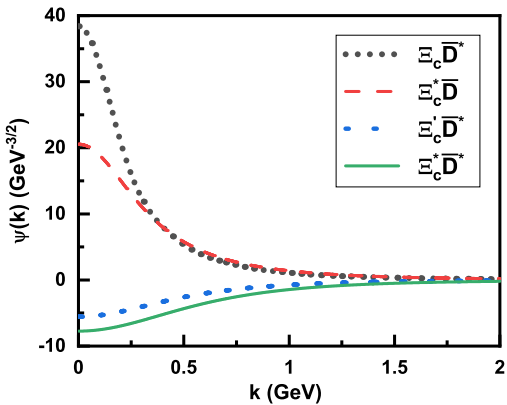}
\caption{The radial wave functions for the $P_{cs}(4459)$ as a coupled $\Xi_c\bar{D}^*/\Xi_c^*\bar{D}/\Xi_c^{\prime}\bar{D}^*/\Xi_c^*\bar{D}^*$ molecule with $I(J^P)=0(3/2^-)$ in the momentum space. It is obtained by performing the Fourier transformation, i.e., $\psi(k)=\int {d^3\bm{r}}e^{i\bm{k}\cdot\bm{r}}\psi(\bm{r})$. Here, its binding energy is $E=-19.28$ MeV.}\label{wave}
\end{figure}

The interaction Lagrangians related to the discussed decay processes are given as \cite{Lin:1999ad,Nagahiro:2008mn,Liu:2001ce,MuellerGroeling:1990cw}
\begin{eqnarray}
\mathcal{L}_{PPV} &=& \frac{iG}{2\sqrt{2}}\left\langle\partial^{\mu}{P}\left({P}{V}_{\mu}-{V}_{\mu}{P}\right)\right\rangle,\\
\mathcal{L}_{VVP} &=& \frac{G'}{\sqrt{2}}\epsilon^{\mu\nu\alpha\beta}
    \left\langle\partial_{\mu}{V}_{\nu}\partial_{\alpha}{V}_{\beta}{P}\right\rangle,\\
\mathcal{L}_{VVV} &=& \frac{iG}{2\sqrt{2}}
    \left\langle\partial^{\mu}{V}^{\nu}\left({V}_{\mu}{V}_{\nu}-{V}_{\nu}{V}_{\mu}\right)\right\rangle,\\
\mathcal{L}_{BBP} &=& g_{p}\left\langle\bar{{B}}i\gamma_5{P}{B}\right\rangle,\label{lag1}\\
\mathcal{L}_{BBV} &=& g_{v}\langle\bar{{B}}\gamma^{\mu}{V}_{\mu}{B}\rangle
              +\frac{f_{v}}{2m}\left\langle\bar{{B}}\sigma^{\mu\nu}\partial_{\mu}{V}_{\nu}{B}\right\rangle,\label{lag2}\\
\mathcal{L}_{BDP} &=& \frac{g_{BDP}}{m_{{P}}}
             \left(\bar{{D}}^{\mu}{B}+\bar{{B}}{{D}}^{\mu}\right)\partial_{\mu}{P},\\
\mathcal{L}_{BDV} &=& -i\frac{g_{BDV}}{m_{{V}}}
        \left(\bar{{D}}^{\mu}\gamma^5\gamma^{\nu}{B}+\bar{{B}}\gamma^5\gamma^{\nu}{{D}}^{\mu}\right)
        \left(\partial_{\mu}{V}_{\nu}-\partial_{\nu}{V}_{\mu}\right).\nonumber\\
\end{eqnarray}
Here, ${P}$, ${V}$, ${B}$, and ${D}$ stand for the pseudoscalar and vector mesons, octet, and decuplet baryons. For example, in the $SU(4)$ quark model, the pseudoscalar and vector mesons are expressed as
\begin{scriptsize}
\begin{eqnarray*}
{P} &=& \left(\begin{array}{cccc}
\frac{\pi^0}{\sqrt{2}}+\frac{\eta}{\sqrt{6}}+\frac{\eta_c}{\sqrt{12}}      &\pi^+     &K^+     &\bar{D}^0\\
\pi^-      &-\frac{\pi^0}{\sqrt{2}}+\frac{\eta}{\sqrt{6}}+\frac{\eta_c}{\sqrt{12}}    &K^0     &D^-\\
K^-     &\bar{K}^0      &-\frac{2\eta}{\sqrt{6}}+\frac{\eta_c}{\sqrt{12}}      &D_s^-\\
D^0      &D^+     &D_s^+    &-\frac{3\eta_c}{\sqrt{12}}\end{array}\right),\\
{V} &=& \left(\begin{array}{cccc}
\frac{\rho^0}{\sqrt{2}}+\frac{\omega_8}{\sqrt{6}}+\frac{J/\psi}{\sqrt{12}}      &\rho^+     &K^{*+}     &\bar{D}^{*0}\\
\rho^-      &-\frac{\rho^0}{\sqrt{2}}+\frac{\omega_8}{\sqrt{6}}+\frac{J/\psi}{\sqrt{12}}    &K^{*0}     &D^{*-}\\
K^{*-}     &\bar{K}^{*0}      &-\frac{2\omega_8}{\sqrt{6}}+\frac{J/\psi}{\sqrt{12}}      &D_s^{*-}\\
D^{*0}      &D^{*+}     &D_s^{*+}    &-\frac{3J/\psi}{\sqrt{12}}\end{array}\right),
\end{eqnarray*}\end{scriptsize}with $\omega_8=\omega \text{cos}\theta+\phi \text{sin}\theta$ and $\text{sin}\theta=-0.761$ \cite{particle}. Coupling constants adopted in the following calculations are estimated from the $\rho-\pi-\pi$, $N-N-\pi$, $N-N-\rho(\omega)$, $N-\Delta-\pi$, and $N-\Delta-\rho$ interactions. For example, when explicitly expanding the the $SU(4)$ invariant interaction Lagrangians between baryons and pseudoscalar mesons, one can obtain
\begin{eqnarray}
\mathcal{L}_{BBP} &=& \frac{5b-4a}{4\sqrt{2}}G_p(\bar{n}i\gamma^5\pi^0n-\bar{p}i\gamma^5\pi^0p\nonumber\\
     &&+\bar{p}i\gamma^5\pi^+n+\bar{n}i\gamma^5\pi^-p)\nonumber\\
     &&+\frac{\sqrt{3}}{8}(a+b)G_p(\bar{\Xi}_c^0i\gamma^5D^0\Lambda^0-\bar{\Xi}_c^+i\gamma^5D^+\Lambda^0)\nonumber\\
     &&+\frac{3}{8}(a-b)G_p(\bar{\Xi}_c^{'0}i\gamma^5D^0\Lambda^0-\bar{\Xi}_c^{'+}i\gamma^5D^+\Lambda^0)\nonumber\\
     &&-\frac{a+b}{4\sqrt{2}}G_p(\bar{\Xi}_c^+i\gamma^5D_s^+\Xi^0+\bar{\Xi}_c^0i\gamma^5D_s^+\Xi^-)\nonumber\\
     &&-\frac{3}{4}\sqrt{\frac{3}{2}}(a-b)G_p(\bar{\Xi}_c^{'+}i\gamma^5D_s^+\Xi^0+\bar{\Xi}_c^{'0}i\gamma^5D_s^+\Xi^-)\nonumber\\
     &&+\frac{\sqrt{3}}{8}(a-2b)G_p(\bar{\Lambda}_c^+i\gamma^5K^0\Xi_c^{+}-\bar{\Lambda}_c^+i\gamma^5K^+\Xi_c^{0})\nonumber\\
     &&-\frac{3}{8}aG_p(\bar{\Lambda}_c^+i\gamma^5K^0\Xi_c^{'+}-\bar{\Lambda}_c^+i\gamma^5K^+\Xi_c^{'0})+\ldots.\label{equation}
\end{eqnarray}
In Refs. \cite{Liu:2001ce,Adelseck:1990ch,Lin:1999ad,Nagahiro:2008mn,Ronchen:2012eg,Janssen:1996kx}, $b/a=5.3$, $g_{\pi NN}=13.5$, $g_{\rho NN}=3.25$, $f_{\rho NN}=6.1$, $g_{BDP}=2.127$, $g_{BDV}=16.03$, $G=12.00$, and $G'=\frac{3 G^2}{\left(32 \sqrt{2}\right) \left(\pi ^2 f_{\pi}^2\right)}$ with $f_{\pi}=0.132$ GeV. All the coupling constants are determined by comparing the corresponding coefficients in Eq. (\ref{equation}). The scattering amplitudes for all the discussed decay processes can be expressed as
\begin{eqnarray}
\mathcal{M}_{{BP\to BV}}^{{P}} &=& g_p\bar{u}_3\gamma_5u_1\frac{1}{q^2-m_{\mathbb{P}}^2}
      {g_{PPV}\epsilon_4^{\mu\dag}(-p_{2\mu}+q_{\mu})},\\
\mathcal{M}_{{BP\to BV}}^{{V}} &=& \left\{g_v\bar{u}_3\gamma^{\mu}u_1+\frac{f_v}{4m^*}\bar{u}_3(\gamma^{\mu}\gamma^{\nu}
      -\gamma^{\nu}\gamma^{\mu})q_{\nu}u_1\right\}\nonumber\\
      &&\times\frac{g_{\mu\beta}-q_{\mu}q_{\beta}/m_{\mathbb{V}}^2}{q^2-m_{\mathbb{V}}^2}
      g_{VVP}\varepsilon^{\lambda\nu\alpha\beta}p_{4\nu}\epsilon_{4\lambda}^{\dag}q_{\alpha},\\
\mathcal{M}_{{BP\to VB}}^{{B}} &=& g_p\bar{u}_4\gamma_5\frac{1}{\rlap\slash{q}-m_{\mathbb{B}}}
     \left\{g_v\epsilon_{3\mu}^{\dag}\gamma^{\mu}u_1\right.\nonumber\\
     &&\left.-\frac{f_v}{4m^*}p_{3\mu}\epsilon_{3\nu}^{\dag}
      (\gamma^{\mu}\gamma^{\nu}-\gamma^{\nu}\gamma^{\mu})q_{\nu}u_1\right\},\\
\mathcal{M}_{{BV\to BV}}^{{P}} &=& g_p\bar{u}_3\gamma_5u_1\frac{-1}{q^2-m_{\mathbb{P}}^2}
      g_{VVP}\varepsilon^{\lambda\sigma\alpha\beta}p_{4\lambda}\epsilon_{4\sigma}^{\dag}p_{2\alpha}\epsilon_{2\beta},\nonumber\\\\
\mathcal{M}_{{BV\to BV}}^{{V}} &=& \left\{g_v\bar{u}_3\gamma^{\mu}u_1
      +\frac{f_v}{4m^*}\bar{u}_3(\gamma^{\mu}\gamma^{\nu}-\gamma^{\nu}\gamma^{\mu})q_{\nu}u_1\right\}\nonumber\\
      &&\times\frac{g_{\mu\beta}-q_{\mu}q_{\beta}/m_{\mathbb{V}}^2}{q^2-m_{\mathbb{V}}^2}
      g_{VVV}\left[\epsilon_{4}^{\alpha\dag}\epsilon_{2}^{\beta}(p_{2\alpha}-q_{\alpha})\right.\nonumber\\
      &&\left.
      -\epsilon_{2\alpha}\epsilon_4^{\alpha\dag}(p_2^{\beta}+p_4^{\beta})
      +\epsilon_{2\alpha}(\epsilon_4^{\beta\dag}q^{\alpha}+p_4^{\alpha}\epsilon_4^{\beta\dag})\right],\\
\mathcal{M}_{{BV\to VB}}^{{B}} &=& \left\{g_v\bar{u}_4\gamma^{\mu}\epsilon_{2\mu}
      +\frac{f_v}{4m^*}\bar{u}_4(\gamma^{\mu}\gamma^{\nu}-\gamma^{\nu}\gamma^{\mu})p_{2\mu}\epsilon_{2\nu}\right\}\nonumber\\
      &&\times\frac{1}{\rlap\slash{q}-m_{\mathbb{B}}}
      \left\{g_v'\gamma^{\alpha}\epsilon_{3\alpha}^{\dag}u_1\right.\nonumber\\
      &&\left.+\frac{f_v'}{4m'}(\gamma^{\alpha}\gamma^{\beta}-\gamma^{\alpha}\gamma^{\beta})p_{3\alpha}\epsilon_{3\beta}^{\dag}u_1\right\},\\
\mathcal{M}_{{DP\to BV}}^{{P}} &=& \frac{g_{BDP}}{m_P}\bar{u}_3q_{\mu}u_1^{\mu}\frac{1}{q^2-m_{\mathbb{P}}^2}
      ig_{PPV}\epsilon_4^{\nu\dag}(q_{\nu}-p_{4\nu}),\\
\mathcal{M}_{{DP\to BV}}^{{V}} &=& -\frac{g_{BDV}}{m_V}\bar{u}_3\gamma^5(\gamma^{\nu}u_{1}^{\mu}-\gamma^{\mu}u_1^{\nu})q_{\mu}\nonumber\\
      &&\times\frac{g_{\nu\beta}-q_{\nu}q_{\beta}/m_{\mathbb{V}}^2}{q^2-m_{\mathbb{V}}^2}
      ig_{VVP}\varepsilon^{\lambda\beta\alpha\delta}q_{\lambda}p_{4\alpha}\epsilon_{4\delta}^{\dag},\\
\mathcal{M}_{{DP\to VB}}^{{B}} &=& ig_{BBP}\bar{u}_4\gamma^5\frac{1}{\rlap\slash{q}-m_{\mathbb{B}}}
      \frac{g_{BDV}}{m_V}\gamma^5(\gamma^{\nu}u_{1}^{\mu}-\gamma^{\mu}u_1^{\nu})q_{\mu}\epsilon_{3\nu}^{\dag},\nonumber\\\\
\mathcal{M}_{{DV\to BV}}^{{P}} &=& \frac{g_{BDP}}{m_P}\bar{u}_3q_{\mu}u_1^{\mu}\frac{i}{q^2-m_{\mathbb{P}}^2}
      g_{VVP}\varepsilon^{\lambda\delta\alpha\beta}p_{2\lambda}\epsilon_{2\delta}p_{4\alpha}\epsilon_{4\beta}^{\dag},\nonumber\\\\
\mathcal{M}_{{DV\to BV}}^{{V}} &=& \frac{g_{BDV}}{m_V}\bar{u}_3\gamma^5(\gamma^{\nu}u_{1}^{\mu}-\gamma^{\mu}u_1^{\nu})q_{\mu}
      \frac{g_{\nu\beta}-q_{\nu}q_{\beta}/m_{\mathbb{V}}^2}{q^2-m_{\mathbb{V}}^2}\nonumber\\
      &&\times g_{VVV}\left[\epsilon_{4}^{\alpha\dag}\epsilon_{2}^{\beta}(p_{2\alpha}-q_{\alpha})\right.\nonumber\\
      &&\left.-\epsilon_{2\alpha}\epsilon_4^{\alpha\dag}(p_2^{\beta}+p_4^{\beta})
      +\epsilon_{2\alpha}(\epsilon_4^{\beta\dag}q^{\alpha}+p_4^{\alpha}\epsilon_4^{\beta\dag})\right],\\
\mathcal{M}_{{DV\to VB}}^{{B}} &=& \left\{g_{BBV}\bar{u}_4\gamma^{\alpha}\epsilon_{2\alpha}
      +\frac{f_{BBV}}{4m^*}\bar{u}_4(\gamma^{\alpha}\gamma^{\beta}-\gamma^{\beta}\gamma^{\alpha})p_{2\alpha}\epsilon_{2\beta}\right\}\nonumber\\
      &&\times\frac{1}{\rlap\slash{q}-m_{\mathbb{B}}}
      \frac{g_{BDV}}{m_V}\gamma^{5}(\gamma^{\nu}u_{1}^{\mu}-\gamma^{\mu}u_1^{\nu})p_{3\mu}\epsilon_{3\nu}^{\dag}.
\end{eqnarray}
Here, $\mathcal{M}_{{i_1i_2\to f_1f_2}}^{E}$ corresponds to the scattering amplitude for the $i_1+i_2\to f_1+f_2$ process by exchanging the hadron $E$. The above scattering amplitudes have the form of
\begin{eqnarray*}
\mathcal{M} &\sim& \frac{c_0+c_1\bm{k}^2+c_2\bm{p}^2+c_3\bm{k}\cdot\bm{p}+c_i(\bm{k}^4, \bm{p}^4, \ldots)}{\bm{k}^2+\bm{p}^2+2\bm{k}\cdot\bm{p}+M^2}.
\end{eqnarray*}
For the heavy loosely bound state, the higher order terms like the $c_i(\bm{k}^4, \bm{p}^4, \ldots)$ contribute very small. In our calculations, we neglect these interactions. According to the relation in Eq. (\ref{amplitude}), the convergence of the amplitude $\mathcal{M}(P_c\to f_1+f_2)$ only depends on the wave functions of the $P_{cs}$ state as shown in Figure \ref{wave}. For simplicity, we set a upper limit integral $k_{\text{Max}}$ on the amplitude $\mathcal{M}(P_c\to f_1+f_2)$ according to the wave function normalization $\int_0^{k_{\text{Max}}} d^3k\psi(\bm{k})^2=1$.

\section{numerical results}\label{sec3}

\begin{figure}[!htbp]
\center
\includegraphics[width=3.in]{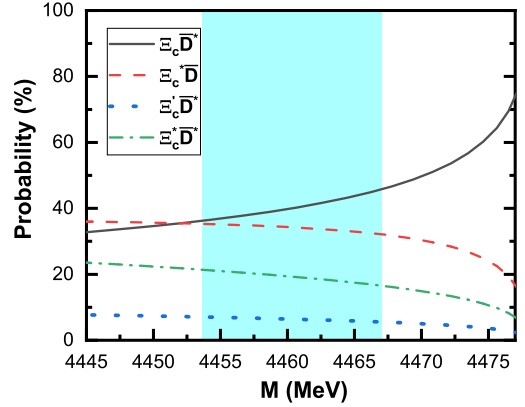}
\caption{Probabilities for different channels of the $P_{cs}(4459)$ as a coupled $\Xi_c\bar{D}^*/\Xi_c^*\bar{D}/\Xi_c^{\prime}\bar{D}^*/\Xi_c^*\bar{D}^*$ molecule with $I(J^P)=0(3/2^-)$. The shallow area labels the position of the $P_{cs}(4459)$ including the experimental uncertainty.}\label{probability}
\end{figure}

Before calculating the decay widths, let's brief introduce the bound state property of the $P_{cs}(4459)$ as a strange hidden-charm meson-baryon molecular pentaquark. In Figure \ref{probability}, we present the probabilities for different channels of the $P_{cs}(4459)$ as a coupled $\Xi_c\bar{D}^*/\Xi_c^*\bar{D}/\Xi_c^{\prime}\bar{D}^*/\Xi_c^*\bar{D}^*$ molecule with $I(J^P)=0(3/2^-)$. Here, the coupled channel effect plays an essential role in forming this bound state. And the $\Xi_c\bar{D}^*$ and $\Xi_c^*\bar{D}$ channels are the most important, followed by the $\Xi_c^*\bar{D}^*$ and $\Xi_c'\bar{D}^*$ channels.

\begin{figure}[!htbp]
\center
\includegraphics[width=3.2in]{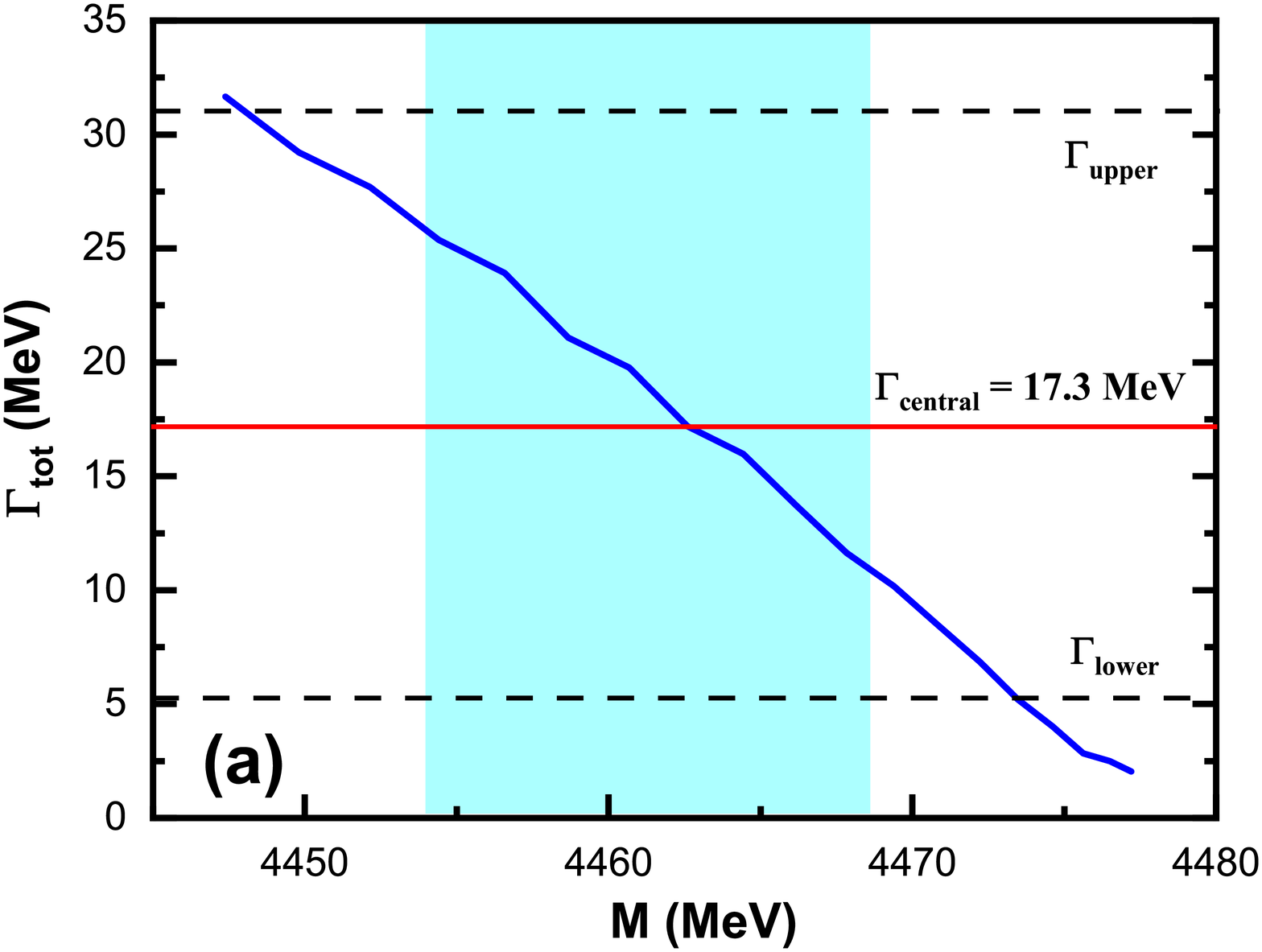}
\includegraphics[width=3.2in]{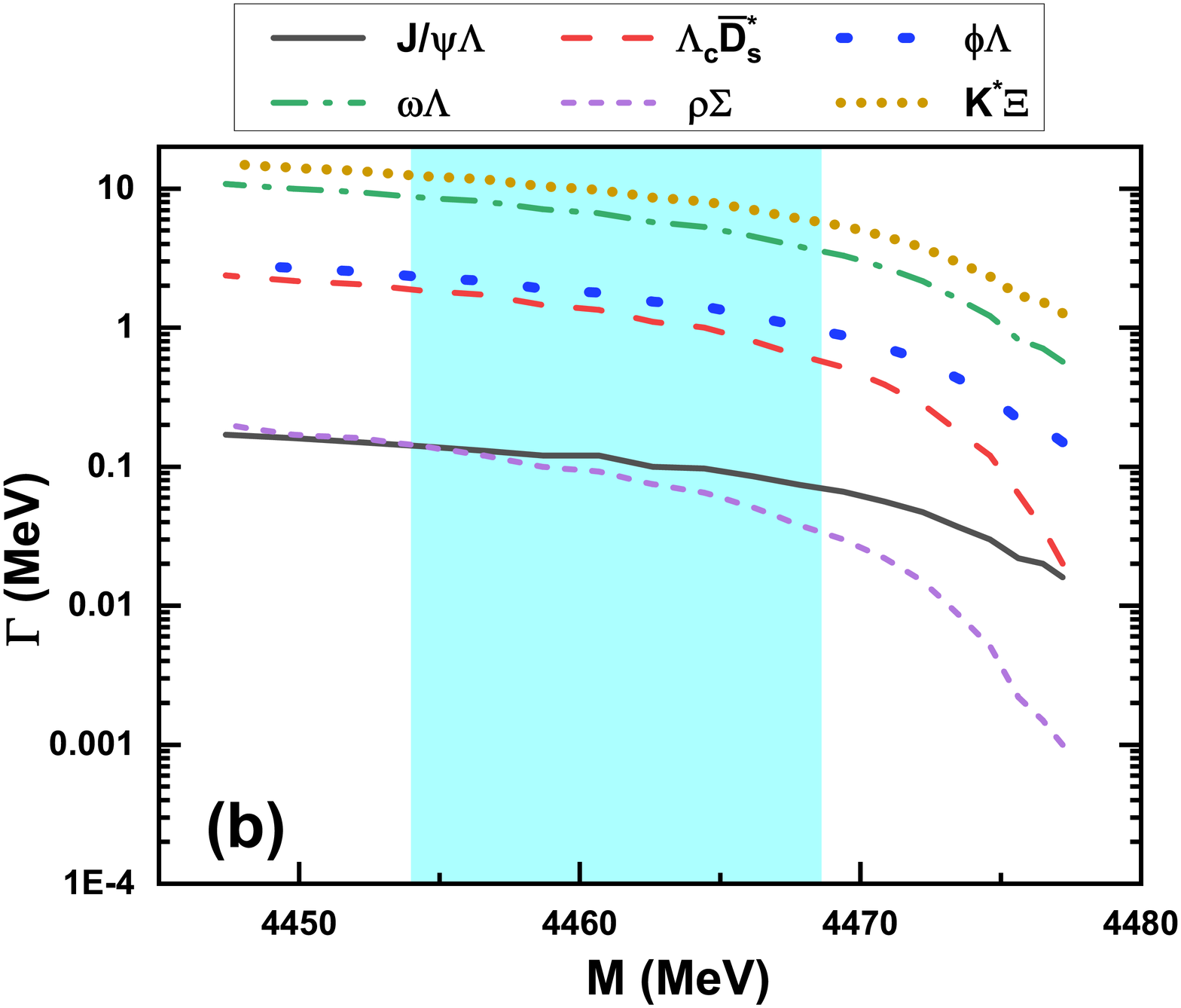}
\caption{The total (a) and partial (b) decay widths for the $P_{cs}(4459)$ as a coupled $\Xi_c\bar{D}^*/\Xi_c^*\bar{D}/\Xi_c^{\prime}\bar{D}^*/\Xi_c^*\bar{D}^*$ molecule with $I(J^P)=0(3/2^-)$. The shallow area labels the position of the $P_{cs}(4459)$ including the experimental uncertainty.}\label{decay}
\end{figure}

With the above preparations, we can further produce the two-body strong decay widths for the coupled $\Xi_c\bar{D}^*/\Xi_c^*\bar{D}/\Xi_c^{\prime}\bar{D}^*/\Xi_c^*\bar{D}^*$ molecule with $I(J^P)=0(3/2^-)$. In Figure \ref{decay}, we present the corresponding decay widths for the $P_{cs}(4459)$. Here, we take the binding energy from $-0.75$ MeV to $-30$ MeV. We see that
\begin{itemize}
  \item The total two-body strong decay width $\Gamma_{\text{tot}}$ is from 10 MeV to 25 MeV in the mass range of the $P_{cs}(4459)$. It is consistent to the experiment data $\Gamma= 17.3\pm6.5_{-5.7}^{+8.0}~\text{MeV}$.

  \item In general, when the hadronic molecular state binds deeper and deeper, the overlap of the wave functions of the components becomes larger and larger. The quark exchange in the hadronic molecular state becomes easier and easier. As shown in Figure \ref{decay}(a), with the decreasing of the mass of the $P_{cs}(4459)$, the total decay width turns larger.

  \item For the $c\bar{c}-$annihilation decay modes, the $K^*\Xi$ and $\omega\Lambda$ decay modes are the most important among all the discussed decay channels as shown in Figure \ref{decay}(b). The partial widths for these two final states are around several or more than ten MeV, the corresponding branch fraction $(\Gamma_{K^*\Xi}+\Gamma_{\omega\Lambda})/\Gamma_{\text{tot}}$ is around 80\%. For the remaining $\phi\Lambda$ and $\rho\Sigma$ channels, their partial decay widths are around a few tenths and percents MeV, respectively.

  \item Compared to the light hadron final states, the hidden-charm decay widths are much smaller as their narrow phase space. Here, the partial decay width for the $P_{cs}(4459)\to J/\psi\Lambda$ process is only several percents MeV.

  \item For the open-charm decay modes, the partial decay width for the $\Lambda_c\bar{D}^{*}$ channel is in the range of 0.6 MeV to 2.0 MeV. The relative ratios for the $\mathcal{R}=\Gamma_{\Lambda_c\bar{D}^{*}}/\Gamma_{J/\psi\Lambda}$ is around ten. Thus, the open-charm decay should be an essential decay mode to search for the $P_{cs}$ state as a strange hidden-charm molecular pentaquark in our model.

\end{itemize}

To summarize, our results of the two-body strong decay widths support the $P_{cs}(4459)$ as the coupled $\Xi_c\bar{D}^*/\Xi_c^*\bar{D}/\Xi_c^{\prime}\bar{D}^*/\Xi_c^*\bar{D}^*$ molecule with $I(J^P)=0(3/2^-)$.

\section{summary}\label{sec4}

In 2019, the LHCb Collaboration reported three narrow hidden-charm pentaquarks ($P_c(4312)$, $P_c(4440)$, and $P_c(4457)$) in the $\Lambda_b\to J/\psi pK$ process \cite{Aaij:2019vzc}. They are likely to be the charmed baryon and anti-charmed meson molecular states. And the coupled channel effect plays an very important role in forming a bound state and the strong decay \cite{Wang:2019spc,Chen:2019asm}. Very recently, the LHCb Collaboration continued to report an evidence of the hidden-charm pentauqarks with strangeness $|S|=1$. After adopting the OBE model and considering the coupled channel effect, we find the newly reported $P_{cs}(4459)$ can be regarded as the coupled $\Xi_c\bar{D}^*/\Xi_c^*\bar{D}/\Xi_c^{\prime}\bar{D}^*/\Xi_c^*\bar{D}^*$ molecule with $I(J^P)=0(3/2^-)$. The dominant channels are the $S-$wave $\Xi_c\bar{D}^*$ and $\Xi_c^*\bar{D}$ channels.

Using the obtained bound wave functions, we study the two-body strong decay behaviors for the $P_{cs}(4459)$ in the molecular picture. Our results show that the total decay width here is well around the experimental value reported by the LHCb Collaboration \cite{1837464}. The $c\bar{c}-$annihilation decay modes are very important. In particular, the partial decay widths for the $P_{cs}(4459)\to K^*\Xi(\omega\Lambda)$ are over several MeV, their branch fractions are nearly 80\%. The partial decay width for the $\Lambda_cD_s^*$ mode is around 1 MeV. The relative ratio for the $\mathcal{R}=\Gamma_{\Lambda_c\bar{D}_c^{*}}/\Gamma_{J/\psi\Lambda}$ is around 10.

Until now, the inner structure and the spin-parity of the $P_{cs}(4459)$ are still mystery, more theoretical and experimental studies are needed. Although our phenomenological study is still model dependence, the strong decay information provided here can be a crucial test of the hadronic molecular state assignment to the $P_{cs}$ state. Experimental search for the possible hidden-charm molecular pentaquark will be helpful to check and develop these adopted phenomenological models.

\section*{ACKNOWLEDGMENTS}

Rui Chen is very grateful to Xiang Liu and Shi-Lin Zhu for helpful discussions and constructive suggestions.
This project is supported by the National Postdoctoral Program for Innovative Talent, the National Natural Science Foundation of China under Grants No. 11705069 and No. 11575008, and National Key
Basic Research Program of China (2015CB856700).

\end{document}